\documentclass[a4paper]{article}
\usepackage[applemac]{inputenc}
\usepackage[english]{babel}
\usepackage{amsmath,latexsym, eucal}
\usepackage{amssymb}
\usepackage{ctable}
\usepackage{authblk}
\usepackage{url}
\usepackage{marvosym}
\usepackage{pgf}
\usepackage{tikz}
\usetikzlibrary{arrows,automata}
\usepackage{graphicx}
\DeclareGraphicsExtensions{.eps,.pdf,.png,.jpg}

\begin{document}

\title{Analysis of cloud storage prices}
\author[1]{Loretta Mastroeni}
\author[2]{Maurizio Naldi}
\affil[1]{Universit\`{a} di Roma Tre, Dipartimento di Economia, Via Silvio D'amico 77, 00145 Rome, Italy}
\affil[2]{Universit\`{a} di Roma Tor Vergata, Dipartimento di Ingegneria Civile ed Ingegneria Informatica, Via del Politecnico 1, 00133 Roma, Italy}
\date{\today}
\maketitle
\tableofcontents

\newpage
\section{Introduction}
Cloud storage is a fast growing service, whereby an individual or a company stores its data on a storage facility owned and managed by a third party (the cloud provider). The actual storage facility may be positioned at a single location or scattered around the globe, but the cloud user does not need to know. 

Cloud users can eliminate their own storage infrastructure, relying on the cloud only. The migration from an owned infrastructure to a leased one has the immediate benefit of avoiding capital investments in favour of a more flexible expense management based on operational expenses only \cite{NaldiNGI11}. The decision to migrate has however to weigh the risk components associated to both solutions.  Capital investments are one-off in nature but they may lead to savings in the long run \cite{NaldiCEC2011}. In addition, switching to the cloud may expose the cloud user to the lock-in phenomenon and price rises (though the decision to continue buying disks is exposed to disk price rises as well) \cite{NaldiWPin}. But a cloud user can also keep its own infrastructure and use the cloud for backup purposes only.

In either case, the cloud storage market is forecast to expand, and the need to evaluate the different offers grows as well. Though a service can be thoroughly described just by the complete set of its features, the most relevant point of differentiation seems to be the price. The risk of commoditization of cloud computing (of which cloud storage may be an ancillary function) has been pointed out by Durkee \cite{Durkee10}, as has been the case with the web hosting industry. In fact, at present cloud providers are pushing price as the most attractive leverage to get users choose their platform.

Though we realize that the complexity of a service proposition cannot be captured by price only, in this paper we focus on price. A comparison of the technical merits of commercial cloud platforms has been reported in \cite{Hu2010}, but no economic analysis has been accomplished so far. However, a comparison of pricing plans for cloud storage is needed to decide whether to migrate or not. In every analysis conducted so far about the opportunity to migrate, one or more pricing plans have been adopted to draw conclusions about that opportunity: in \cite{Walker10} and \cite{NaldiNGI11}, just Amazon's prices have been employed, while in \cite{Abu-LibdehPW10} those prices has been used along with four other providers (Rackspace, GoGrid, Nirvanix, and EMC Atmos). In addition, analysing the pricing plans proposed by cloud providers helps understand the price structure of the industry and its economies of scale.

In this paper, we conduct a survey of pricing plans proposed by major cloud providers. All prices have been gathered on the web and are correct at the time of writing, though they may vary in the future. 

For each pricing plan, we compute the unit price to uncover the strength of economies of scale. In addition, we fit the pricing data to a basic semi-variable price model, which includes just two parameters: the fixed fee and the variable price per unit. We perform a Pareto-dominance analysis and find that some pricing plans are dominated by others and could be removed from the shortlist of providers to consider.

The paper is structured as follows. In Section \ref{market}, we provide some indications about the current market size and composition. For the major providers we provide all the pricing data in Section \ref{obsprice}. The pricing plans are then classified and compared in Section \ref{Pricomp}. We finally introduce the two-part tariff approximation and use it to perform a Pareto-dominance analysis in Section \ref{twoapp}. 

\section{The market of cloud storage}
\label{market}
The cloud storage market is rapidly increasing. A number of providers offer solutions both for consumers and companies, with the list increasing each day (Google is the latest addition with its Google Drive service). In this section, we provide some figures for the market size and list the major cloud providers we are going to examine in this paper.

Official figures for the present size of the cloud storage market are not available. Some estimates have been distributed, typically by consultant firms. IDC estimates that the total spending on storage systems, software, and professional services by public cloud service providers will increase to \$10.9 billion in 2015 \cite{IDC}. An alternative estimate by the Taneja Group gives a figure of \$4B for the market today, with a growth to almost \$10B by 2014 \cite{Boles}, giving a compound annual growth rate of over 35\%. According to a recent study released by Gartner, the percentage of consumer content stored in the cloud will grow from a slight 7\% in 2011 to 36\% in 2016 \cite{Verma}.

We have collected data from the major cloud providers. We base our analysis on publicly advertised prices and don't consider those companies that, though offering cloud storage services, do not provide a public price list or provide storage just as a part of an inclusive service. In the following sections, we report the results for the following providers:
\begin{itemize}
\item Dropbox;
\item Crashplan;
\item SugarSync;
\item IDrive;
\item Google Drive;
\item Box;
\item Mozy;
\item Carbonite;
\item Symform;
\item Amazon.
\end{itemize}

\section{Survey of advertised prices}
\label{obsprice}
For our survey we have considered a wide range of cloud providers, as detailed in Section \ref{market}. We have collected pricing information on their websites and obtained the unit prices. In this section, we provide the details of the price survey for each provider. All prices are correct at the time of writing, though they may vary in the future. Though prices may be originally given in \$, we have converted all the money amounts in euros through a fixed conversion rate (1 \EUR = 1.3 \$). In order to get a level comparison, all prices are referred to a month of usage. 

Many providers address separately the consumer and the business markets, by providing specific pricing plans. Different prices for the two categories come along with differentiated service features, among which the most relevant seems to be the number of computers that can be backed up on the cloud platform. Typically, consumer pricing plans allow for just one computer, while business pricing plans all cater for more users. We stick to the consumer vs business classification.

Whenever a provider offers several pricing packages which address the same category, in the following we consider the cheapest one for the amount of memory required.

\subsection{Dropbox}
Dropbox (\url{www.dropbox.com}) offers a very popular storage service, which features both a consumer and a business category. A free basic service is available for consumers, with a storage capacity up to 2 GB, though the maximum capacity may be increased up to 32 GB by inviting friends to join Dropbox (with 500 MB of additional space gained for each joining friend). In the paid packages, the maximum capacity envisaged is 100 GB for consumers, while no limit is advertised for business customers. 

\begin{figure}[htbp]
 \begin{minipage}[b]{5.5cm}
   \centering
   \includegraphics[width=5.5cm]{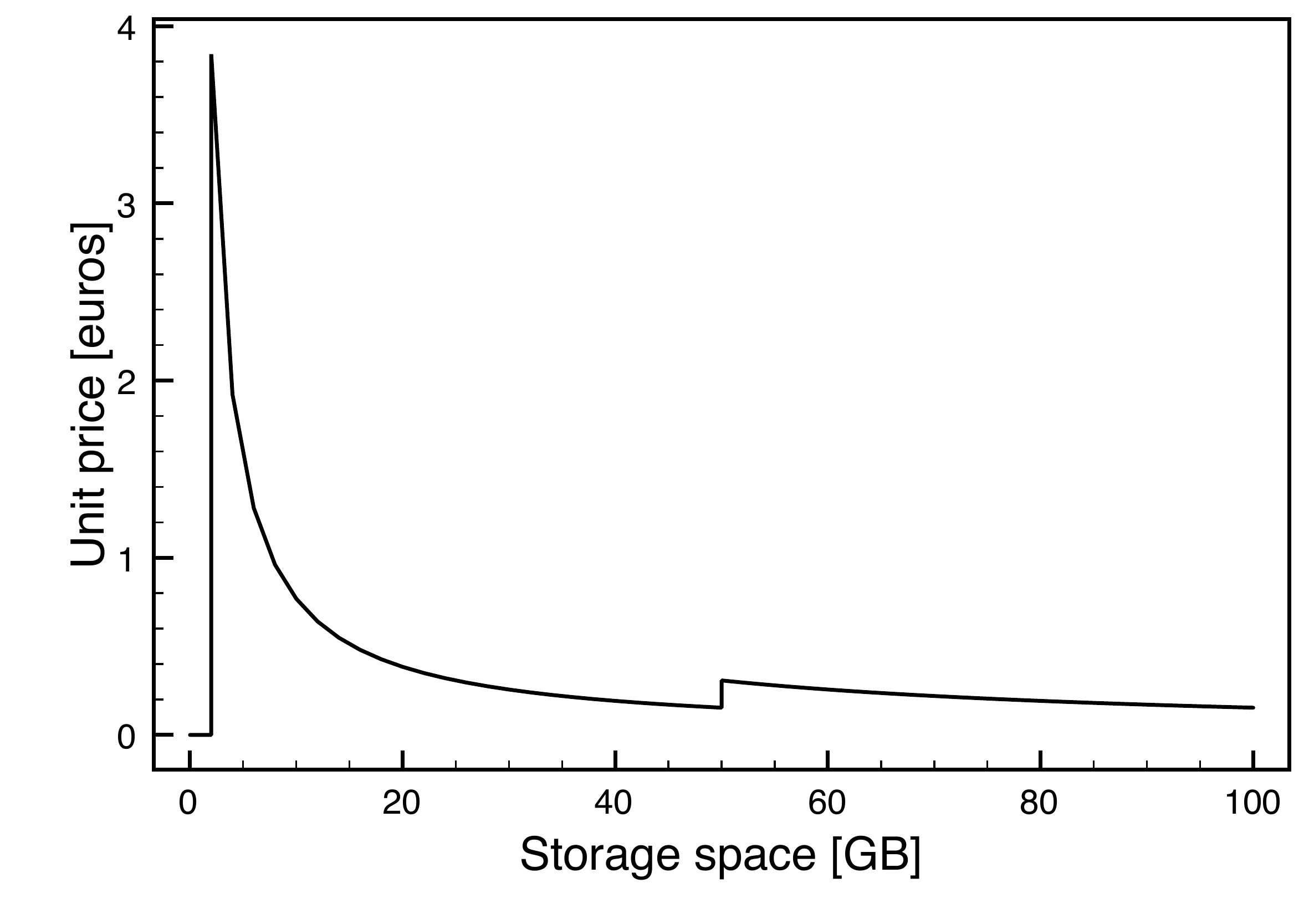}
\caption{Unit prices of Dropbox for consumers}
\label{fig:dropc}
 \end{minipage}
 \ \hspace{2mm} \hspace{3mm} \
 \begin{minipage}[b]{5.5cm}
  \centering
   \includegraphics[width=5.5cm]{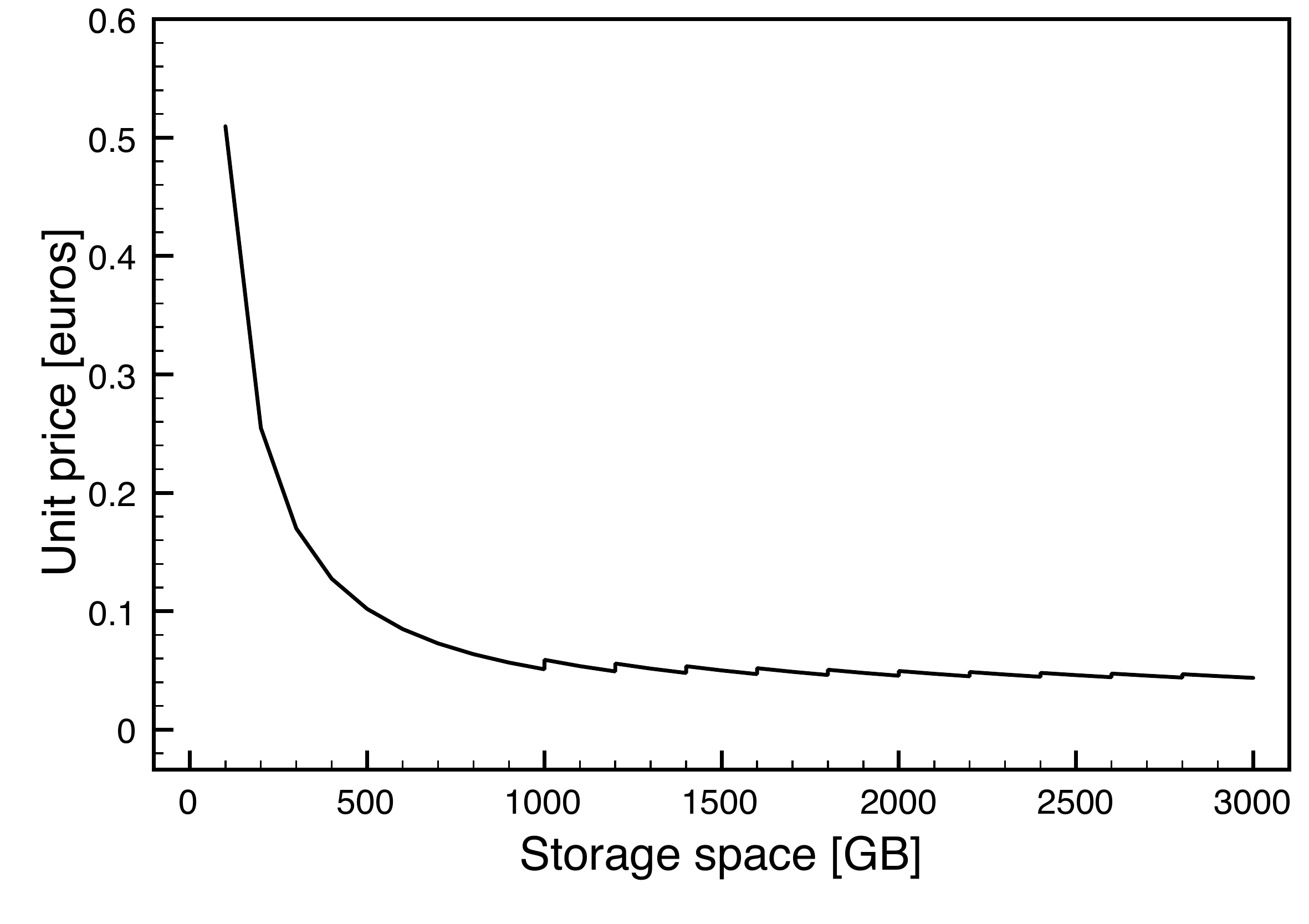}
\caption{Unit prices of Dropbox for business customers}
\label{fig:dropb}
 \end{minipage}
\end{figure}

For consumers, just two paid packages are offered (\textit{Pro50} and \textit{Pro100}), which charge respectively 9.99 \$/month and 19.99 \$/month. A referral bonus is available for these schemes also, with 1 GB of additional space per referral. The resulting monthly unit prices are shown in \figurename~\ref{fig:dropc}, where we have neglected the possible bonus due to referrals. If we neglect the portion of the curve corresponding to the free package, the price curve is piecewise hyperbolic, with two local minima, corresponding respectively to the passage from \textit{Pro50} to \textit{Pro100} and to the full exploitation of the \textit{Pro100} package. By joining the local minima, we obtain a baseline providing the unit prices of full capacity exploitation for each pricing package. In this case, the two local minima give roughly the same unit price of 0.154 \EUR/month.

A pricing formula based on the number of customers is instead implied for business customers. In the business package (named \textit{Teams}), the number of users is employed as the price driver, rather than the amount of memory. However, a relation is provided between the number of users and the maximum amount of available memory. The basic offer considers 5 users, to which 1 TB is associated, and then adds 200 GB for each additional user. The basic package is priced at 795 \$ per year, and each additional chunk of 200 GB comes at 125\$ per year. On the basis of Dropbox pricing information, we can derive the following formula relating the monthly price $P$ (in \$) to the number $k$ of customers 
\begin{equation}
\label{unitpdrop}
P=\frac{170+125 k}{12} \qquad k\ge 5.
\end{equation}
The local minima of the monthly unit price, obtained when each capacity chunk is fully exploited, are given by the following formula
\begin{equation}
p_{k}=\frac{170+125 k}{2400 k} \qquad k\ge 5,
\end{equation}
which is a decreasing function of $k$ (made of hyperbolic sections), starting at 0.06625 \$ (0.051 \EUR) and tending to the limit value 0.052 \$, roughly equivalent to 0.04 euros, just 26\% of the minimum price available to consumers. 

The monthly unit price for business customers is shown in \figurename~\ref{fig:dropb}. Again, we observe the same sawtooth-like trend and the overall economy of scale.

\subsection{Crashplan}
Crashplan offers four plans, named respectively \textit{Crashplan}, \textit{Crashplan+}, \textit{Crashplanpro}, and \textit{Crashplanpro+}, which are advertised on its site \url{www.crashplan.com}. 

Among the four plans, the first one is a free package explicitly aimed at consumers. However, it doesn't actually provide online storage, but rather a backup facility to multiple destinations. The Crashplan+ plan comes instead in three flavours, named respectively \textit{Crashplan+10GB}, \textit{Crashplan+Unlimited}, and \textit{Crashplan+FamilyUnlimited}. All of them are not free, the first two ones being limited to 1 computer (hence, implicitly directed to consumers), and the third one allowing up to 10 computers (catering for families' needs, as its name hints). In return for a fixed fee, Crashplan+10GB allows for 10 GB of online storage, while the other two schemes boast unlimited capacity. The unit price of Crashplan+10GB reaches then a minimum of 0.208\$ per GB per month (0.16 \EUR) when the capacity is fully exploited. As to other schemes, they cannot be compared, since they boast unlimited capacity (which would bring the unit price to zero), a claim which should however be thoroughly verified. They come at a fixed fee  respectively of \$ 3/month/computer (Crashplan+Unlimited) and \$ 6/month/computer (Crashplan+FamilyUnlimited).

\begin{figure}[htbp]
\centering
\includegraphics[width=0.6\columnwidth]{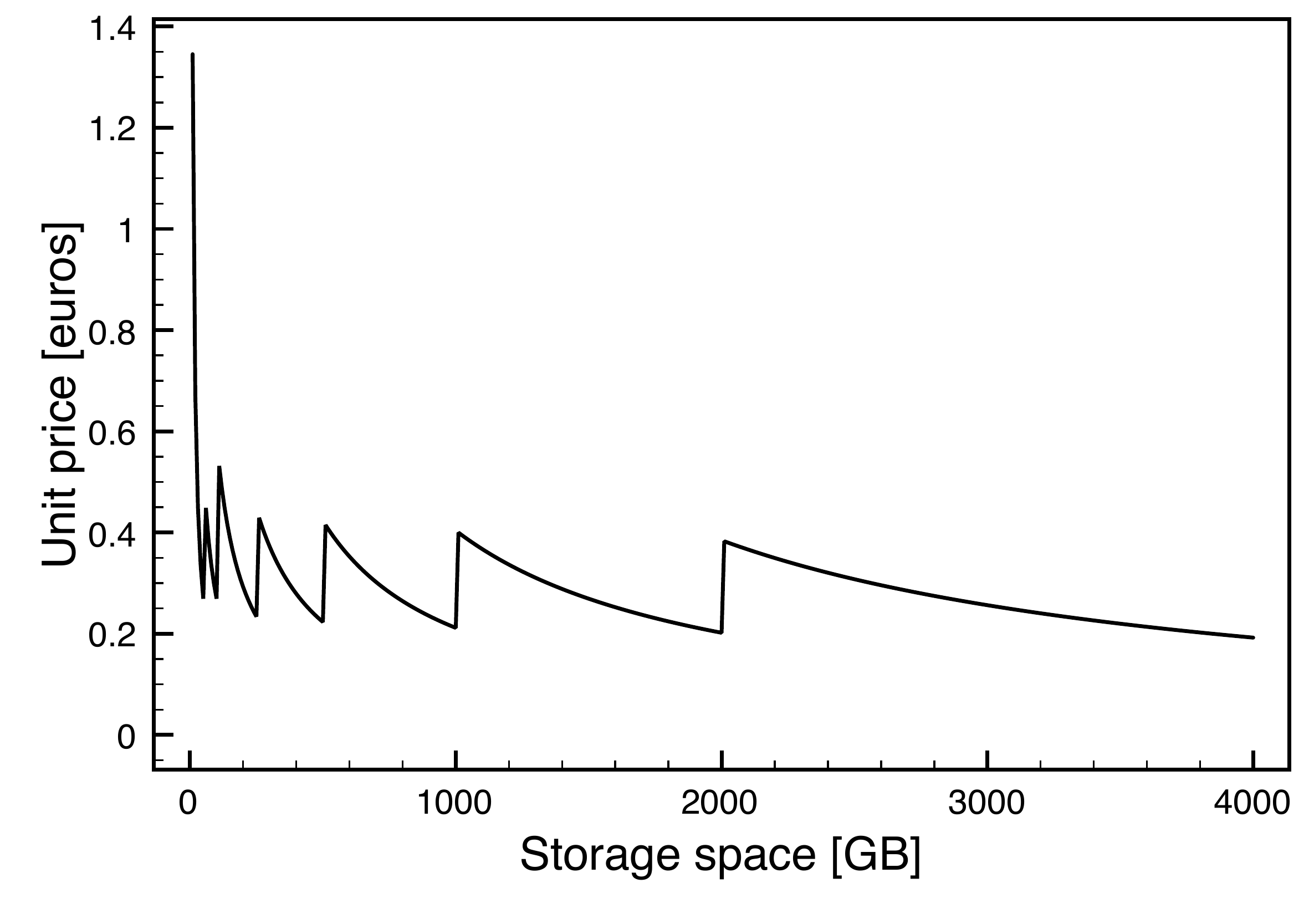}
\caption{Unit prices of Crashplan for business customers}
\label{fig:crash}
\end{figure}

The Crashplanpro package starts instead by offering 50 GB (which are associated to 3 users) and proceeds by adding chunks of capacity as the number of users grows. The resulting unit price is shown in \figurename~\ref{fig:crash}. Despite the usual sawtooth-like graph we can locate a baseline price at 0.2 \EUR, which is 20\% larger than the minimum price established for consumers in the Crashplan+10GB plan. 

\subsection{SugarSync}
\begin{figure}[htbp]
 \begin{minipage}[b]{5.5cm}
   \centering
   \includegraphics[width=5.5cm]{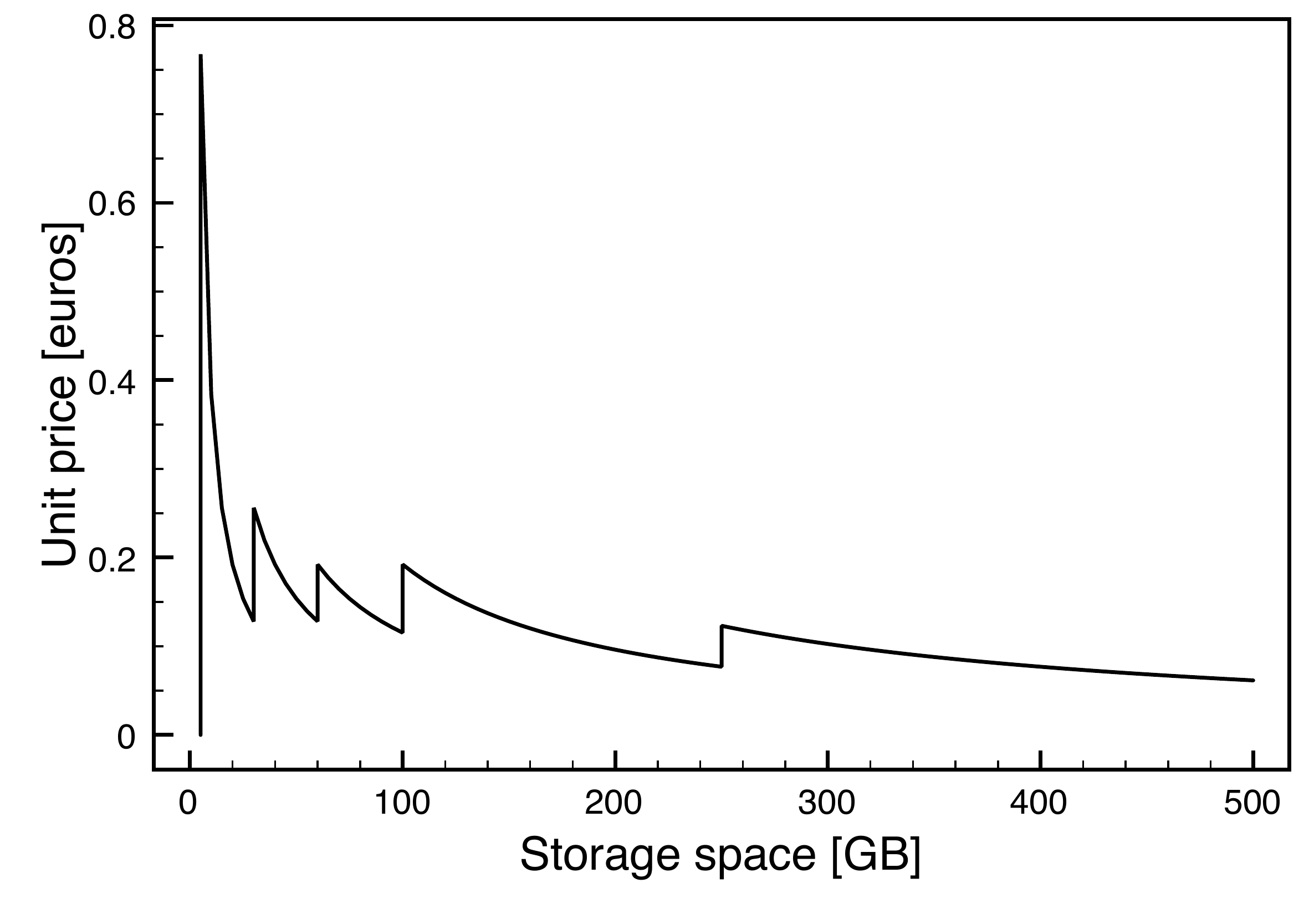}
\caption{Unit prices of SugarSync (consumers)}
\label{fig:sugar}
 \end{minipage}
 \ \hspace{2mm} \hspace{3mm} \
 \begin{minipage}[b]{5.5cm}
  \centering
   \includegraphics[width=5.5cm]{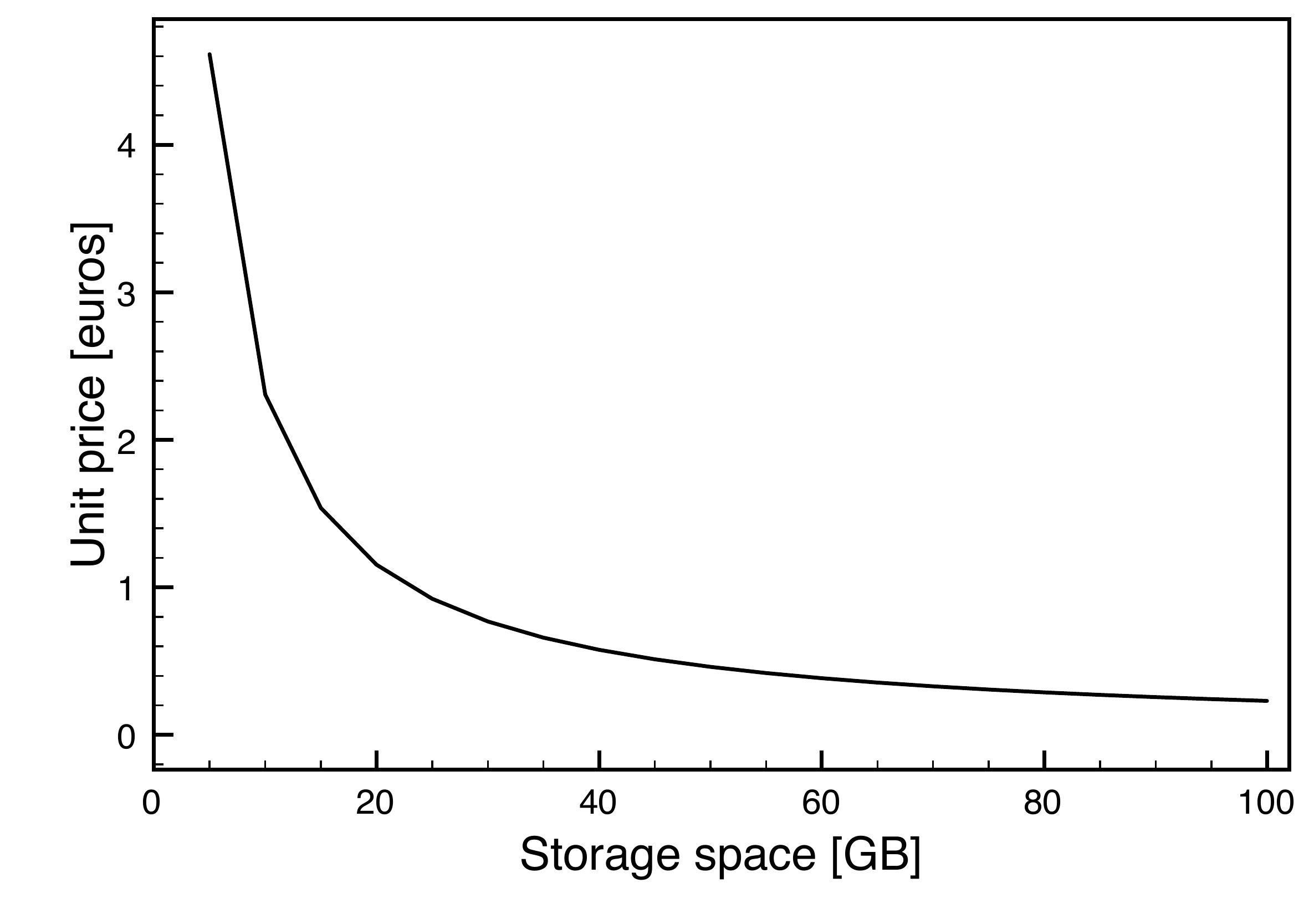}
\caption{Unit prices of SugarSync (business)}
\label{fig:sugarbus}
 \end{minipage}
\end{figure}

SugarSync (\url{www.sugarsync.com}) proposes several plans directed to consumers, starting with a free plan with a maximum capacity of 5 GB. The paid plans consider increasing capacity brackets, up to the maximum capacity of 500 GB. The resulting unit prices are shown in \figurename~\ref{fig:sugar}, where we can observe the usual sawtooth-like curve. Here the imaginary baseline is slightly downward, from 0.128 \EUR~at the first breakpoint at 30 GB down to 0.061 \EUR~at the maximum capacity.

For business customers SugarSync advertises a single plan (others can be provided on demand, but details are not publicly available), whereby a maximum capacity of 100 GB is offered at a fixed fee of 29.99 \$/month. The resulting unit price is shown in \figurename~\ref{fig:sugarbus}.

\subsection{IDrive}
IDRive (\url{www.idrive.com}) offers online storage services both for consumers and business customers. Though they adopt a classification based on the \textit{Basic} and \textit{Pro} categories (made respectively of one and three packages), we find more suitable to reclassify them according to our categories related to the nature of the customers:
\begin{itemize}
\item Basic (consumer);
\item Pro Personal (consumer);
\item Pro Family (consumer);
\item Pro Business (business).
\end{itemize}

The three packages directed to consumers are easily distinct, since they provide different storage limits. In particular the Basic package is free but offer no more than 5 GB of storage space.

\begin{figure}[htbp]
\centering
\includegraphics[width=0.6\columnwidth]{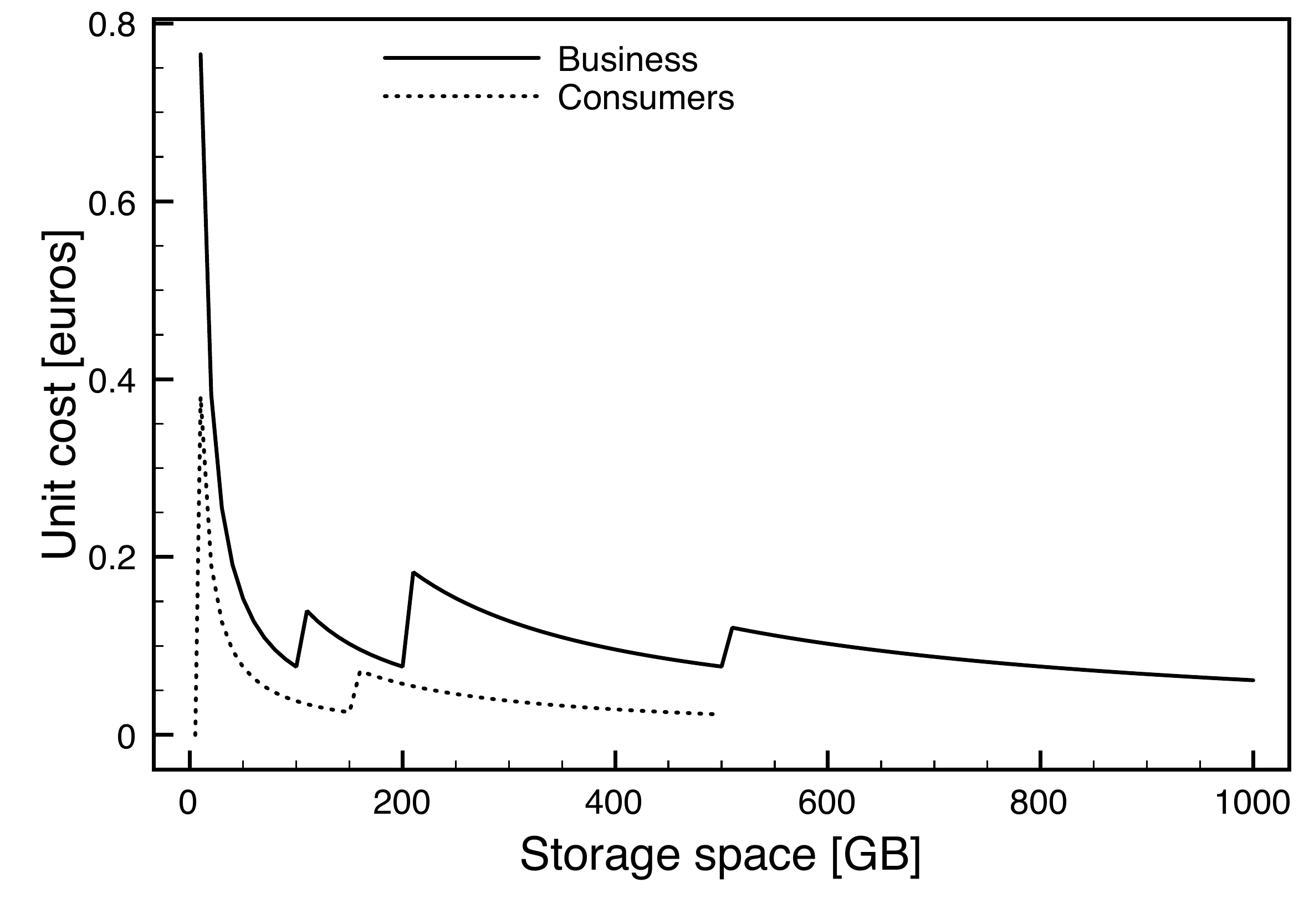}
\caption{Unit prices of IDrive}
\label{fig:idrive}
\end{figure}

In \figurename~\ref{fig:idrive}, we plot the unit prices for both categories. Again, we observe that prices for business customers are higher, with a baseline value that is slightly more than 0.06 euros, nearly three times as much as the minimum price offered to consumers (0.023 euros when the capacity of the Pro Family package is fully exploited).

\subsection{Google Drive}
Google has very recently launched its cloud storage service, named \textit{Google Drive}, whose details are available at \url{drive.google.com}.

Google Drive include a free plan, which allows users to upload up to 5 GB. The free plan is associated to other Google services, which actually increase the amount of info that can be stored on Google's servers. In addition to the 5 GB stored on Google Drive, users can use 10 GB on Gmail and 1 GB on Picasa for free.

\begin{figure}[htbp]
\centering
\includegraphics[width=0.6\columnwidth]{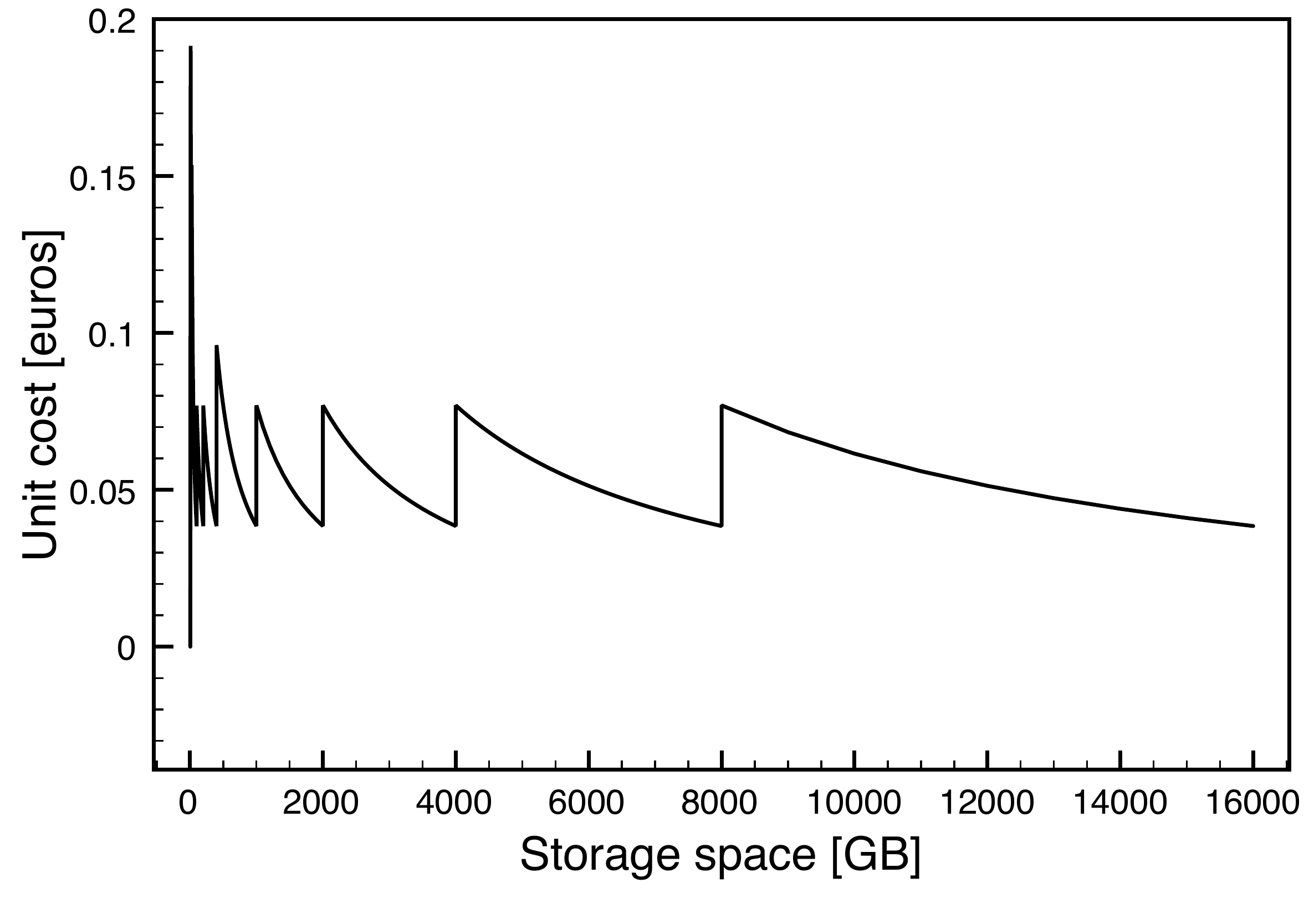}
\caption{Unit prices of Google Drive}
\label{fig:gdrive}
\end{figure}

There appears to be a single set of pricing plans, which are not explicitly directed to either business customers or consumers. However, the absence of particular features in the service advertisement leads to classify that set under the consumer label. The offer is proposed in the usual form of a set of fixed fees, each associated to a maximum capacity. The resulting unit price is shown in \figurename~\ref{fig:gdrive}. We find the usual sawtooth-like curve, with a remarkable baseline at 0.05 \$ (roughly equivalent to 0.0384 \EUR).

\subsection{Carbonite}
\begin{figure}[htbp]
\centering
\includegraphics[width=0.6\columnwidth]{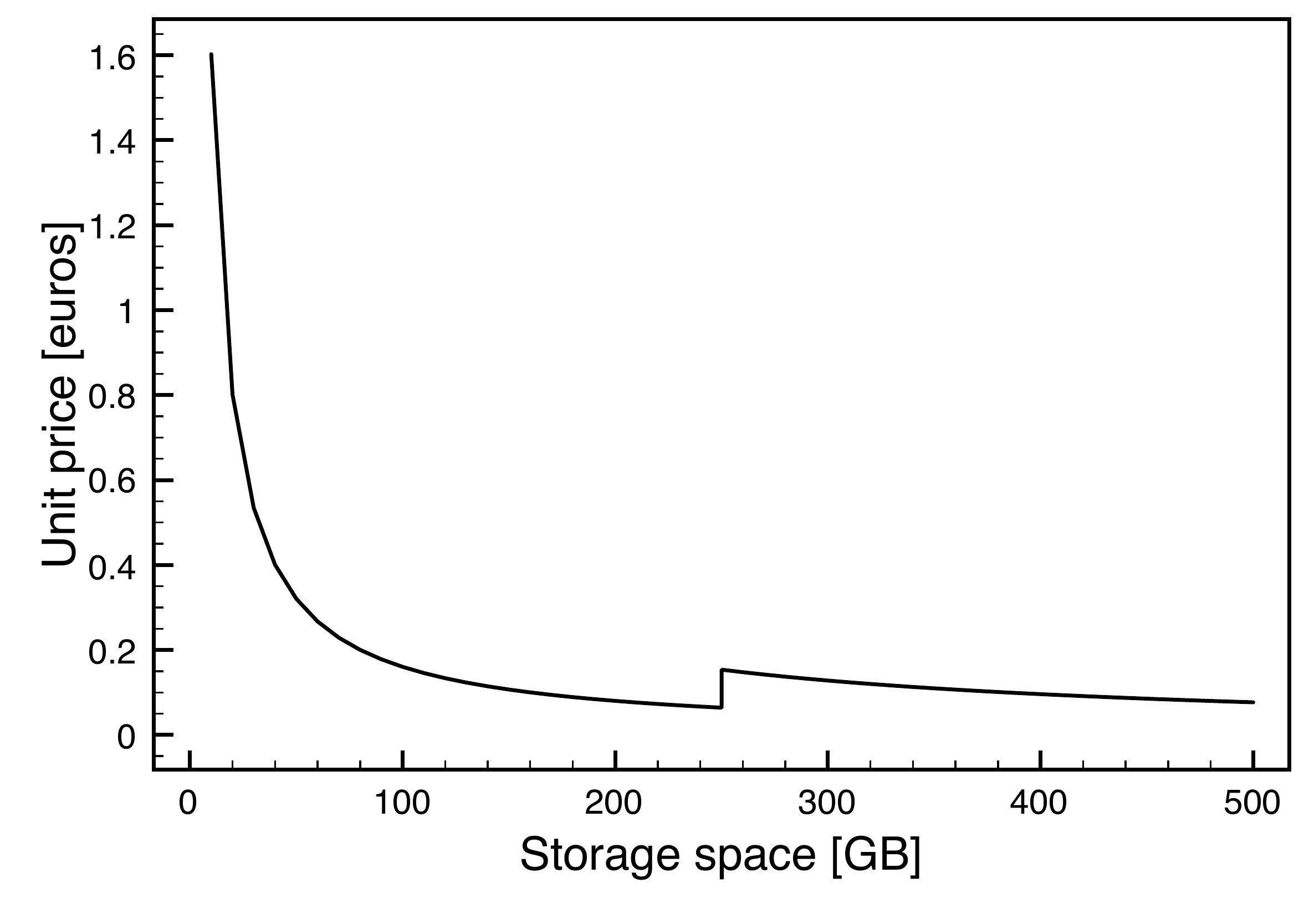}
\caption{Unit prices of Carbonite for business customers}
\label{fig:carbo}
\end{figure}

Carbonite (\url{www.carbonite.com}) offers pricing plans for both customer categories. There are three pricing plans for consumers, which all boast an unlimited online backup for a fixed fee. The fee  is different for the three pricing plans (named \textit{Home}, \textit{HomePlus}, and \textit{HomePremier}), reflecting different features (see Table \ref{table:carbo}). The Home plan does not include External hard drive backup and Mirror Image backup. The HomePremier plan includes also the Courier Recovery service. As to business customers, Carbonite offers instead two pricing plans, which are differentiated by storage capacity: the \textit{Business} plan accepts up to 250 GB; the \textit{BusinessPremier} accepts up to 500 GB. By considering the two plans as a single offer subdivided into two capacity brackets, we obtain the unit price shown in \figurename~\ref{fig:carbo}, with the usual step  in the passage from a bracket to the next.
\begin{table}
\begin{center}
\begin{tabular}{lc}
\toprule
Pricing plan & Yearly fee [\$]\\
\midrule
Home & 59\\
HomePlus & 99\\
HomePremier & 149\\
\bottomrule
\end{tabular}
\caption{Pricing plans of Carbonite for consumers}
\label{table:carbo}
\end{center}
\end{table}

\subsection{Symform}
\begin{figure}[htbp]
\centering
\includegraphics[width=0.6\columnwidth]{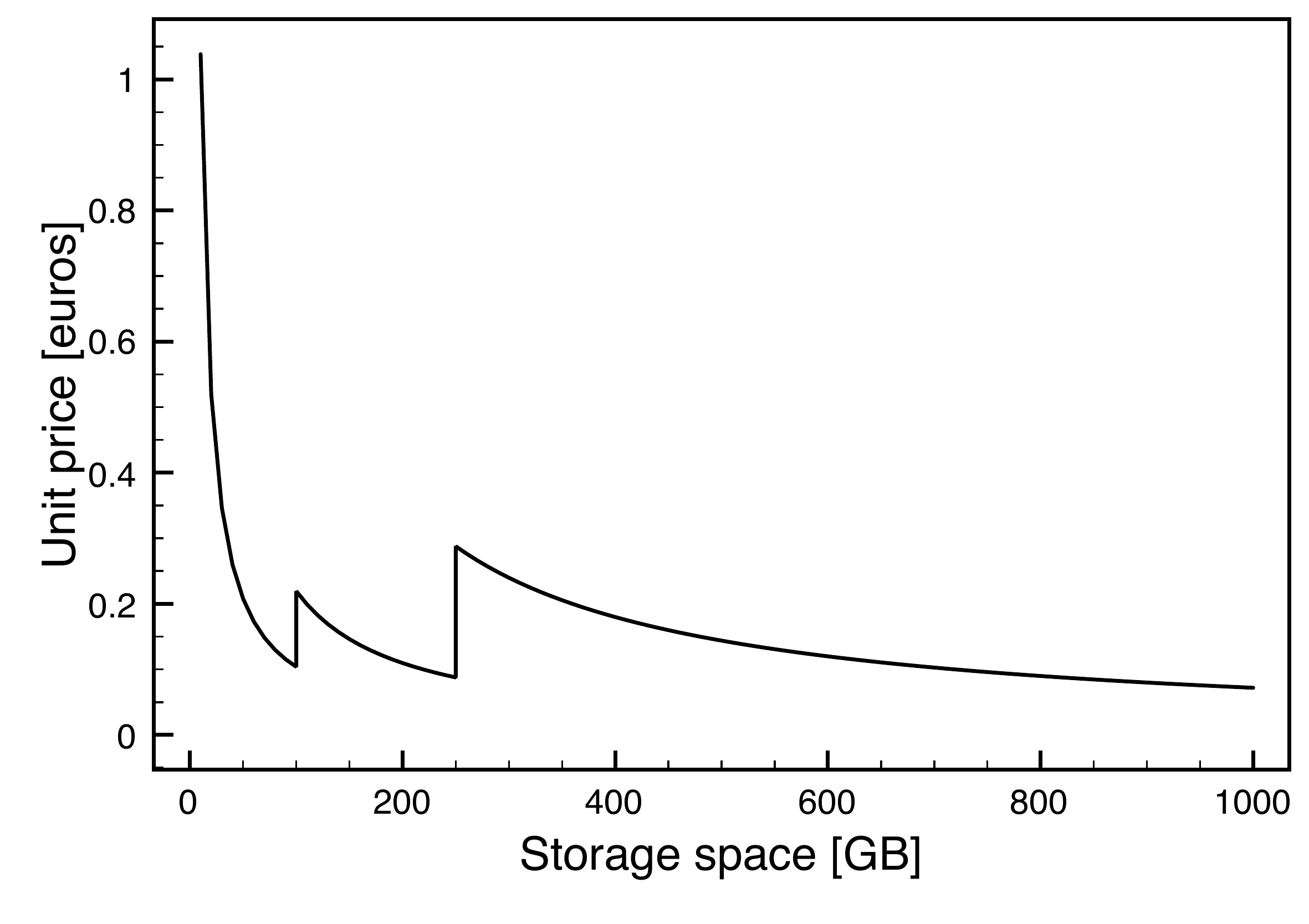}
\caption{Unit prices of Symform}
\label{fig:sym}
\end{figure}

The solution offered by Symform (\url{www.symform.com}) is different from what most cloud providers propose. Symform describes itself as a network, where each member of the network contributes excess local storage in exchange for affordable cloud storage. This peer-to-peer network builds therefore a mutual backup system. Aside from security and privacy considerations, the peculiarity of Symform's solution makes it noncomparable with the pricing plans we have analysed so far. However, Symform allows also for members of its network not contributing storage space, by paying a license fee and a larger storage fee. The license fee is 3.5 \$ per end-user and 50 \$ per server. The storage capacity is subdivided into three brackets, up to 1 TB. We can build a pricing plan similar to those offered to consumers by other providers by considering a single end-user. The resulting unit price is shown in \figurename~\ref{fig:sym}. The shape is quite similar to what we have observed for other providers.
 
\subsection{Mozy}
\begin{figure}[htbp]
 \begin{minipage}[b]{5.5cm}
   \centering
   \includegraphics[width=5.5cm]{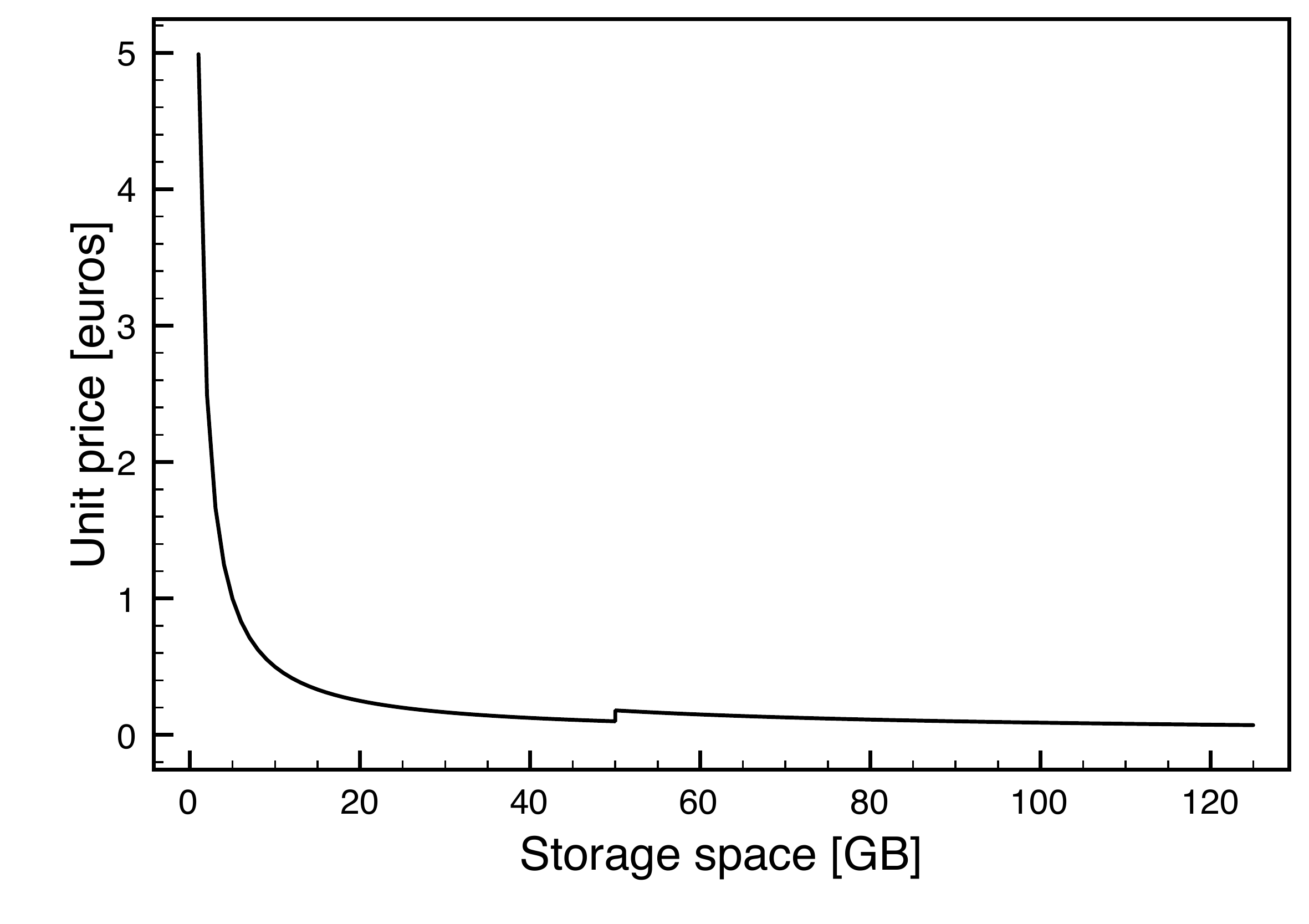}
\caption{Unit prices of Mozy for consumers}
\label{fig:mozyh}
 \end{minipage}
 \ \hspace{2mm} \hspace{3mm} \
 \begin{minipage}[b]{5.5cm}
  \centering
   \includegraphics[width=5.5cm]{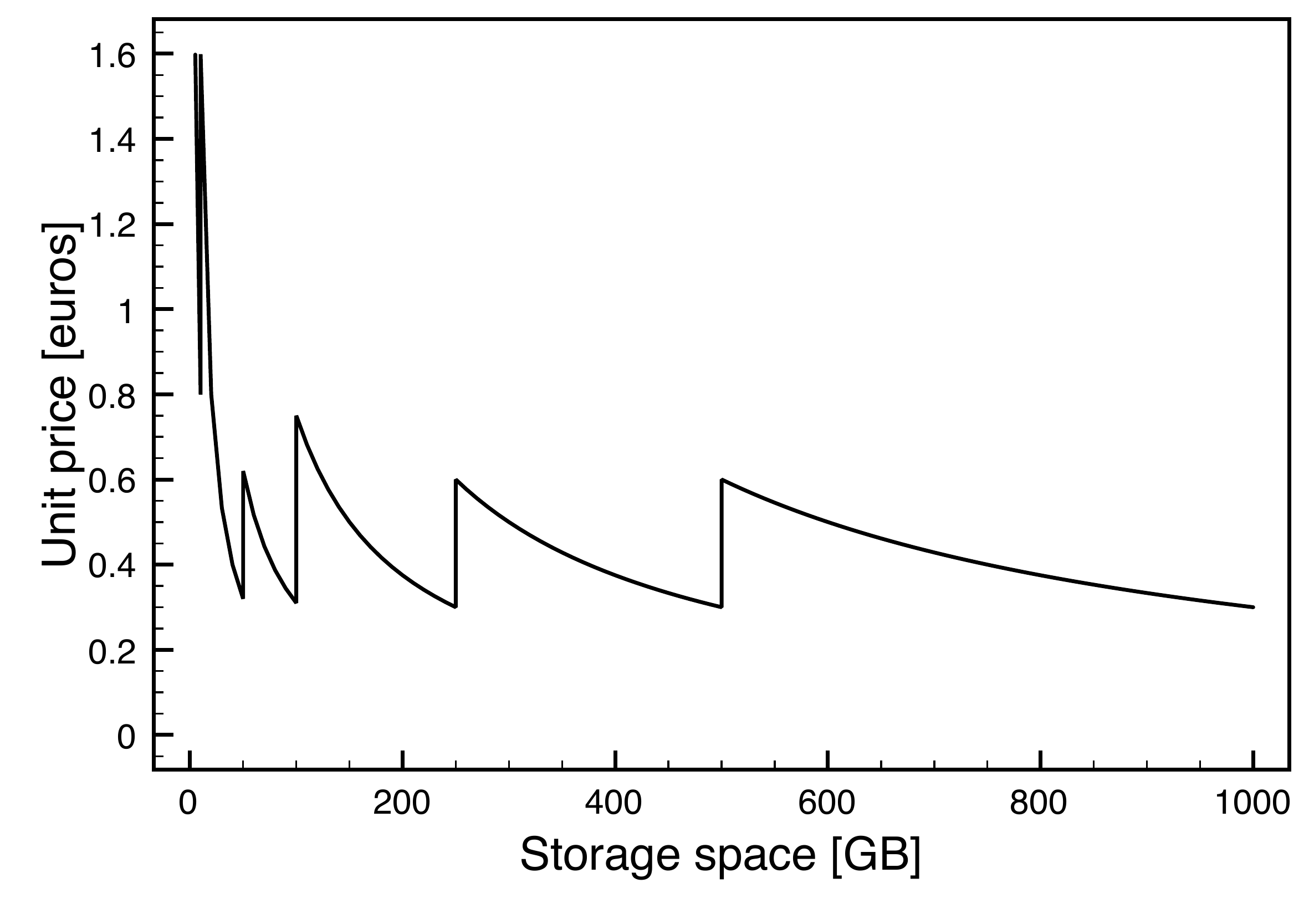}
\caption{Unit prices of Mozy for business customers}
\label{fig:mozyb}
 \end{minipage}
\end{figure}

Mozy (\url{mozy.ie}) has pricing plans for both customer categories, plus an offer directed at enterprises (with prices available on-demand, which we have then neglected in our survey). The offer for consumers is divided into two brackets, with the second one allowing for three computers rather than just one. The limit capacity is 125 GB for consumers and 1 TB for business customers. The resulting unit price is shown in \figurename~\ref{fig:mozyh} and \figurename~\ref{fig:mozyb} respectively. In the business case we can locate a baseline at 0.3 \EUR.

\subsection{Amazon}
Amazon proposes a storage service named \textit{Simple Storage Service} (often identified as Amazon S3). Details on the service can be obtained on their website \url{aws.amazon.com/s3/}. Their offer includes two versions: Standard Storage and Reduced Redundancy Storage. The latter trades reduced reliability for a discounted price. 

The price list is shown in Table \ref{table:pricesamaz}. It is to be noted that Amazon provides unit prices, while all the other providers release lump prices valid up to a specified amount of data. The discount obtained by reducing reliability ranges from 23\% to 33\%. In addition, we notice that the range of data volumes supported by the service is very wide, with a maximum capacity of 5 million GB..

\begin{table}
\begin{center}
\begin{tabular}{ccc}
\toprule
Maximum amount of data [TB] & \multicolumn{2}{c}{Price per month per GB[\$]}\\
& Standard & Reduced Redundancy\\
\midrule
1 & 0.125 & 0.093\\
50 & 0.110 & 0.083\\
500 & 0.095 & 0.073\\
100 & 0.090 & 0.063\\
5000 & 0.080 & 0.053\\
10000 & 0.055 & 0.037\\
\bottomrule
\end{tabular}
\caption{Price list of Amazon}
\label{table:pricesamaz}
\end{center}
\end{table}

\begin{figure}
\begin{center}	
  \includegraphics[width=.6\columnwidth]{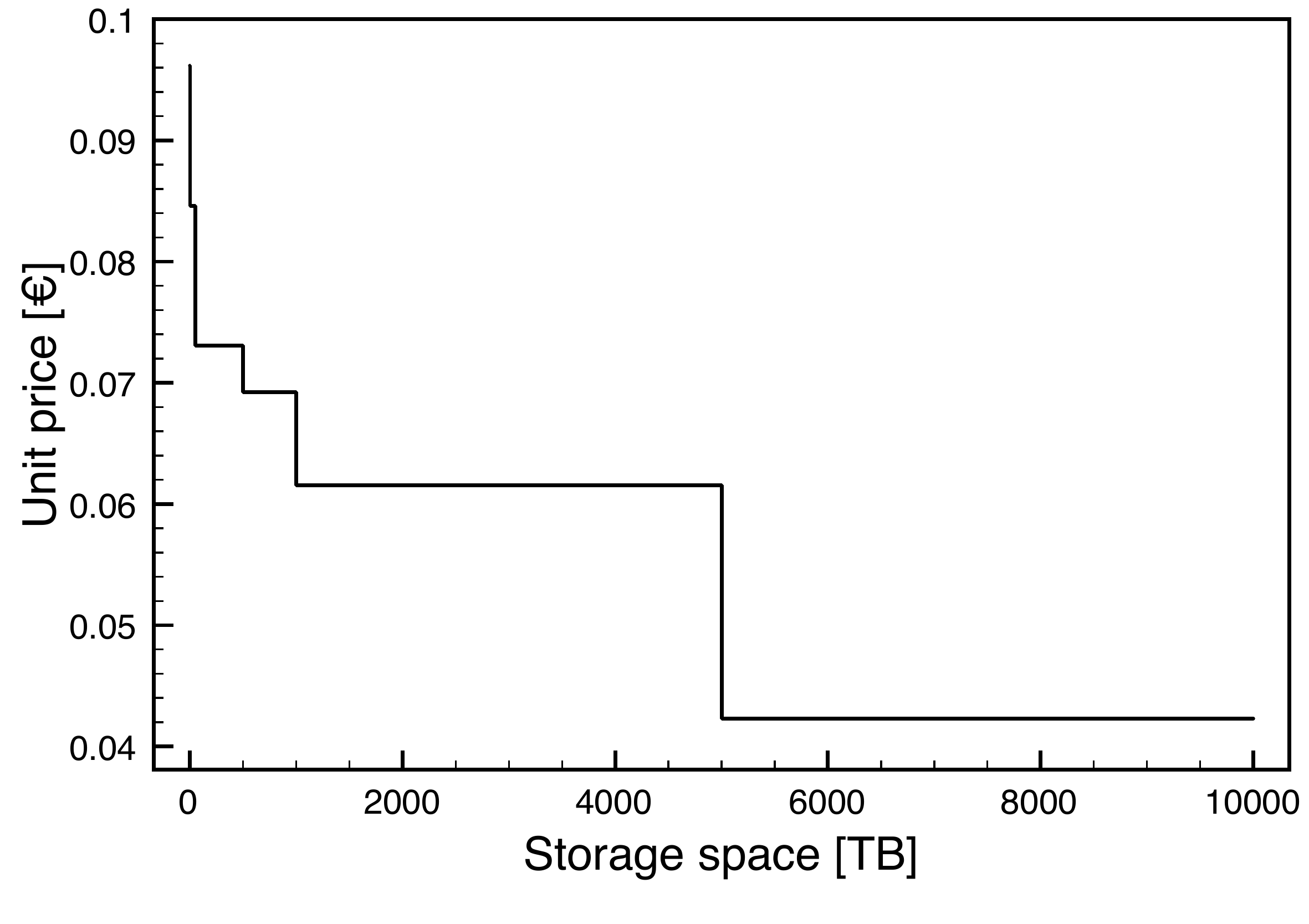}
	\caption{Unit price of Amazon S3}
	\label{fig:pamaz-e}
\end{center}
\end{figure} 

In the following, we consider just the standard version. The resulting unit price curve (a stairwise one) is shown in \figurename~\ref{fig:pamaz-e}.	We observe a significant volume discount.

\section{Comparing and classifying pricing plans}
\label{Pricomp}
In Section \ref{obsprice}, we have reported the pricing plans of a wide selection of  cloud providers. In this section, we proceed to categorize those plans and compare them, searching for the cheapest ones.

We start by observing that most pricing plans allow for some free space (see Table \ref{table:free}). The typical value is 5 GB, which, though a very small fraction of the typical storage space on one's own laptop computer (of the order of one or few percentage points), could be enough for storing data that a customer wants to share temporarily.  
\begin{table}
\begin{center}
\begin{tabular}{lc}
\toprule
Provider & Free storage space [GB]\\
\midrule
Dropbox & 2\\
SugarSync & 5\\
IDrive & 5\\
Google Drive & 5\\
\bottomrule
\end{tabular}
\caption{Amount of free space}
\label{table:free}
\end{center}
\end{table}

Aside from the free storage offered for limited quantities, all the pricing plans currently adopted by cloud providers allow for price discrimination by quantity. However, they are not all equal and fall into two categories: declining block rate charge and bundling.

The first pricing model is also named taper and is a particular type of the general block rate pricing, where the range of consumption is subdivided into subranges and the unit price is held constant over each subrange. More formally, in a block rate tariff the overall price charged to the customer for a volume of consumption $x$ is
\begin{equation}
\label{blockprice}
p = \left\{ \begin{array}{ll}
v^{(1)}x & \textrm{if $0<x\le q^{(1)}$}\\ v^{(1)}q^{(1)}+v^{(2)}(x-q^{(1)}) & \textrm{if $q^{(1)}<x\le q^{(2)}$}\\ \cdots \\\sum_{i=1}^{m-1}v^{(i)}q^{(i)}+v^{(m)}(x-q^{(m-1)}) & \textrm{if $q^{(m-1)}<x\le q^{(m)}$ }
\end{array} \right.
\end{equation}
where the $v^{(i)}$s are the sequence of marginal prices, and the $q^{(i)}$s bracket the subranges over which the marginal price is held constant. In Equation (\ref{blockprice}), we assume that the cloud provider does not provide more than $q^{(m)}$ units of storage ($m\ge 2$). In turn, block-rate pricing can be seen as a special form of multi-part tariff, where the fixed fee has been set equal to zero.

\begin{figure}
\begin{center}	
  \includegraphics[width=.6\columnwidth]{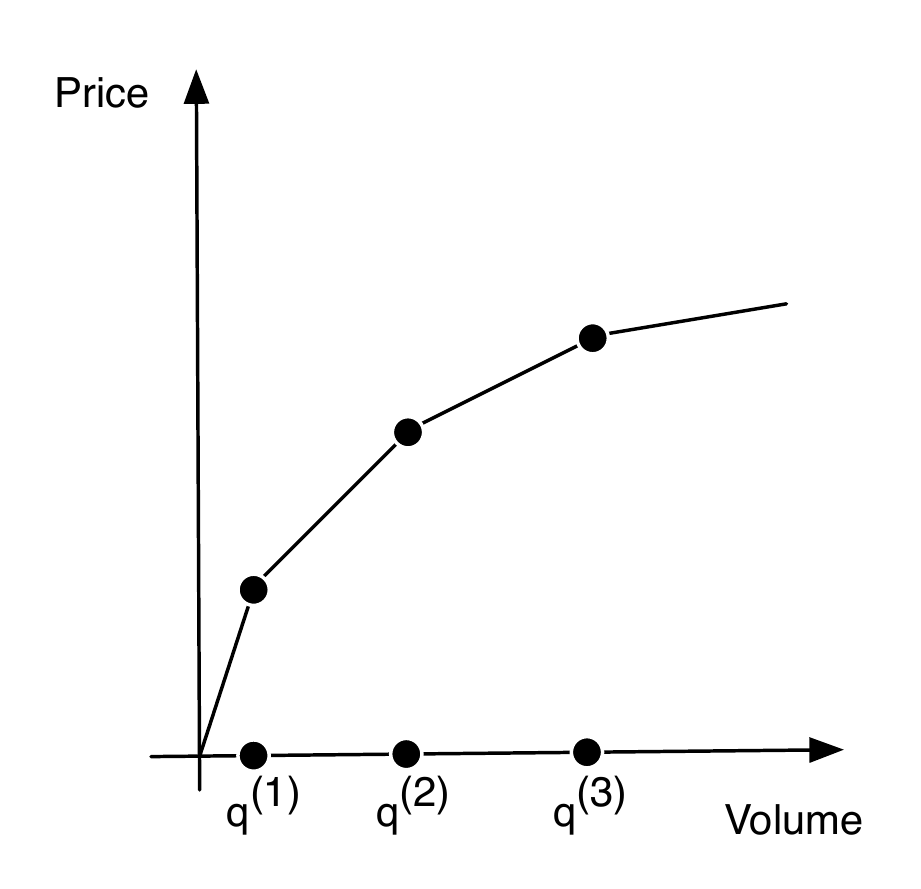}
	\caption{Price-volume relationship in Block rate pricing}
	\label{fig:brate}
\end{center}
\end{figure}

The overall charge is then a piecewise linear function of the amount of storage capacity (see \figurename~\ref{fig:brate}). Diminishing prices at the margin stimulate consumption, which in turn permits the construction of large scale capacity.  

Block rate pricing has been studied by a number of authors. A consistent theory for block tariffs has been developed in \cite{Gabor55}.

Of all the cloud providers examined, just Amazon follows a declining block rate pricing model.

All the other providers adopt instead a bundling pricing model (which in the literature is also called quantity discount). According to the definition given by Shy \cite{Shy2008price}, a seller practices bundling if the firm sells packages containing at least two units of the same product or service, where the price of a package containing several units of the same good is lower than the sum of the prices if the goods were purchased separately. In our case, all the providers opt for multiple bundling, where more than one package is offered for sale, and at least one package contains at least two units.

By using multiple packages, the cloud provider can address the different demand functions of its customers and extract as much profit as possible from each of them. By selecting the right quantities to be included in the bundles, the seller can design a preference revealing mechanism and segment the market among the consumer types. In \cite{Shy2008price} an example is shown with two consumer types and two packages.

Given the dicotomy in the pricing models adopted by cloud providers (though the two groups are extremely unbalanced), it is natural to ask which is better. Kolay and Shaffer have derived the conditions under which bundling is better than block rates for the seller \cite{Kolay03}. For the case of two consumer types that differ in their quantity demanded of a product of fixed quality, they have shown that profits for the seller are higher with bundling than with two-part tariffs as long as standard single-crossing conditions apply and as long as it is optimal to serve all consumers at least some output. The single-crossing condition requires that the demand curve of a high-demand consumer is weakly above the demand curve of a low-demand consumer. Of course, the direction of comparison reverses if we adopt the viewpoint of the customer rather than the cloud provider.

Though the cloud providers, with the exception of Amazon, all adopt the bundling pricing model, the differences among them lie in the price they charge. If we stick to price ad the dominant criterion of choice, the prospective customer has to compare all the proposed pricing plans and pick up the cheapest. Since, the unit price varies widely with the amount of storage leased, in the following we perform a pointwise comparison, by identifying the cheapest cloud provider for each capacity bracket. We underline that the comparison is conducted by taking into account just prices, without reference to the additional features that the pricing plan may incorporate.

We report the result in Table \ref{table:comp}. As to consumers, the field is largely dominated by Google Drive (for smaller capacity values) and IDrive (for moderate to large capacity values). In the 30-100 GB range the unit price of both providers are practically identical, differing for less than 0.001 \EUR. For business customers, the dominant players for small to moderate capacity values (roughly up to 500 GB) are Amazon and again IDrive, while for larger capacity values Dropbox takes definitely the lead. 
\begin{table}
\begin{center}
\begin{tabular}{cc}
\toprule
Capacity bracket [GB] & Most convenient provider\\
\midrule
\multicolumn{2}{c}{Consumers}\\
10-25 & Google Drive\\
30-100  & Google Drive and IDrive\\
110-500 & IDrive\\
\midrule
\multicolumn{2}{c}{Business}\\
10-70 & Amazon\\
80-100 & IDrive\\
110-150 & Amazon\\
160-200 & IDrive\\
210-400 & Amazon\\
410-500 & IDrive\\
510-520 & Amazon\\
530-10000 & Dropbox\\
\bottomrule
\end{tabular}
\caption{Comparison of unit prices}
\label{table:comp}
\end{center}
\end{table}

\section{A two-part tariff approximation and Pareto dominance}
\label{twoapp}
In Section \ref{Pricomp}, we have classified the pricing plans and compared them on a pointwise basis. Though all but one belong to the class of bundling packages, the pricing plans differ from one another for the choice of prices and bundles, which makes a pointwise comparison the only possibility. However, such a comparison does not tell us anything about the structural properties of the pricing plan. We recognize that bundling is a form of nonlinear pricing, whereby the price changes with quantity to reflect the presence of fixed costs and the variation of marginal costs. For the purpose of comparing the structure of pricing plans, we can consider the simplest form of nonlinear pricing: a two-part tariff. In this section, we develop a two-part tariff approximation for all the pricing plans considered and use that approximation to classify them. 

\begin{figure}[htbp]
 \begin{minipage}[b]{5.5cm}
   \centering
   \includegraphics[width=5.5cm]{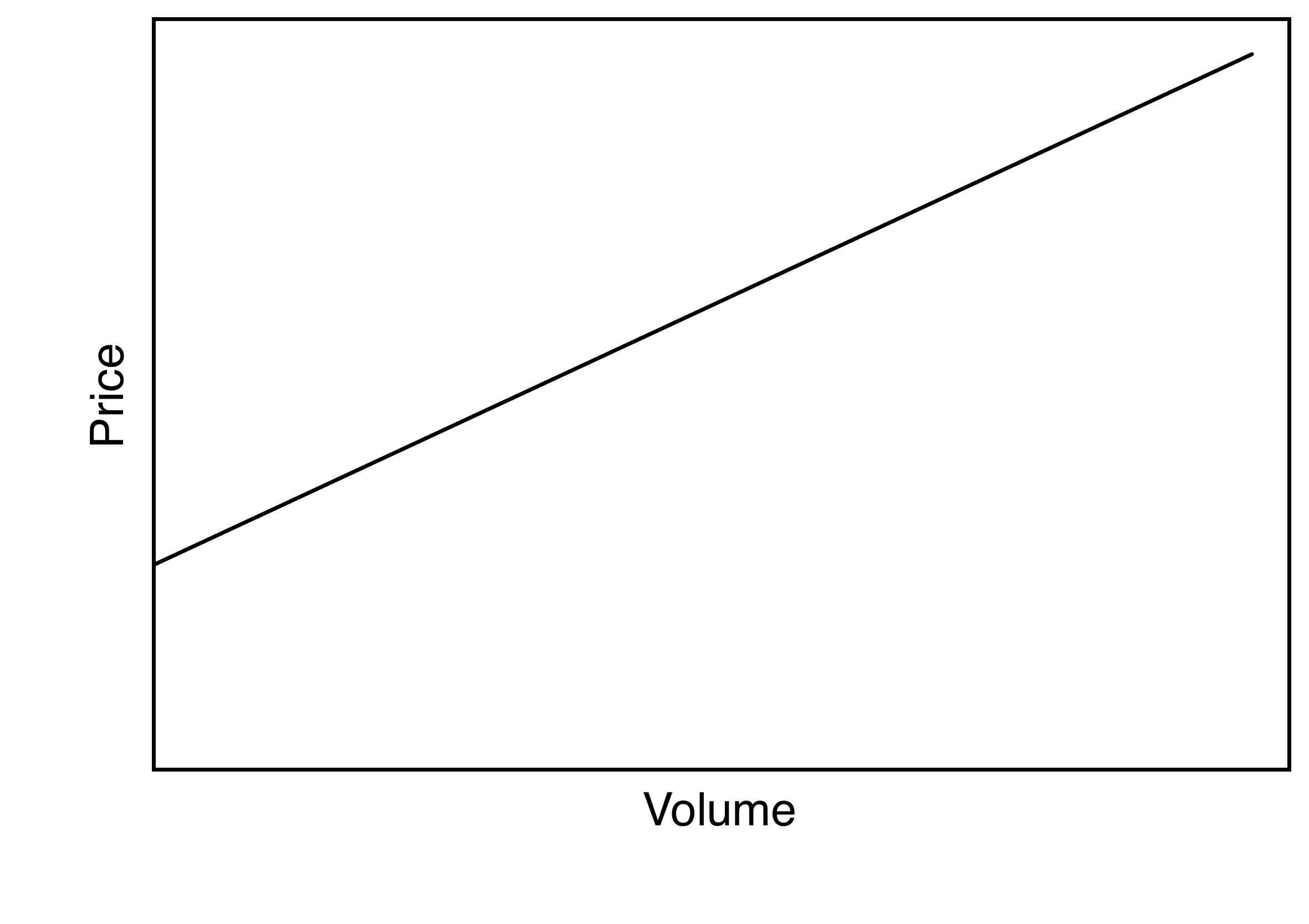}
   \caption{Price in a two-part tariff}
   \label{fig:ptot}
 \end{minipage}
 \ \hspace{2mm} \hspace{3mm} \
 \begin{minipage}[b]{5.5cm}
  \centering
   \includegraphics[width=5.5cm]{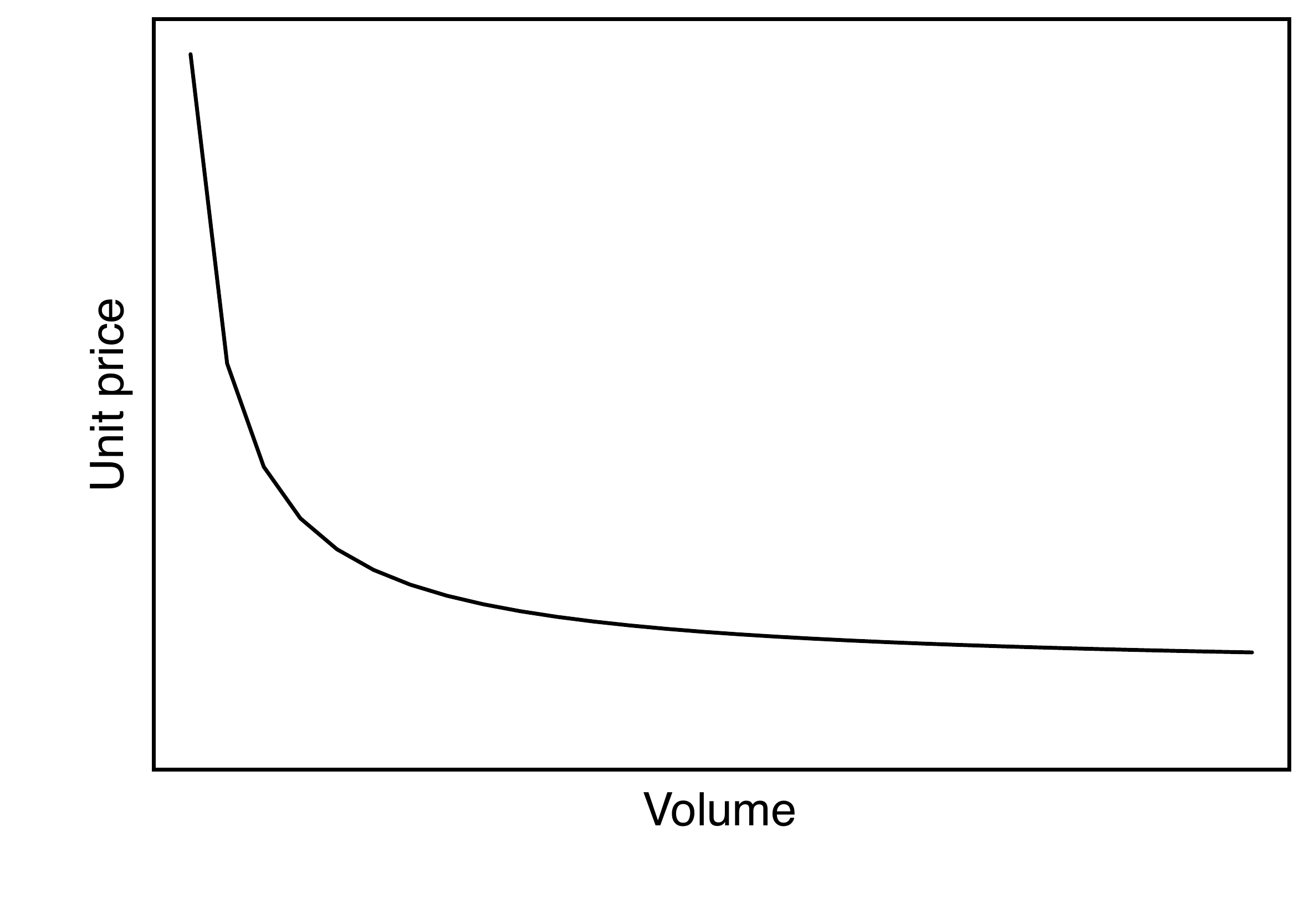}
   \caption{Unit price in a two-part tariff}
   \label{fig:punit}
 \end{minipage}
\end{figure}

In the two-part tariff scheme, the customer pays an initial fixed fee $f$ for the first unit (often justified as a subscription, access, or installation charge), plus a smaller constant price for each unit \cite{Wilson1997}. The overall price charged to the customer is
\begin{equation}
\label{totcost}
p^{(x)}=f+v\cdot x,
\end{equation}
where $v$ is the marginal price, and $x$ is the volume of consumption, i.e., the amount of storage capacity. The resulting amount charged to the consumer is a linear function of the storage capacity (see \figurename~\ref{fig:ptot}). For large volumes the fixed fee is gradually absorbed and its impact is less relevant, highlighting the economy of scale embedded in the service process. The unit price is  
\begin{equation}
\label{unitcost}
p^{(1)}=\frac{f}{x}+v,
\end{equation}
which has the shape of a hyperbola as in \figurename~\ref{fig:punit} and asymptotically tends to the marginal price $v$.

In order to obtain a two-part approximation for the bundling pricing plans shown in Section \ref{obsprice}, we must estimate the values of the two parameters $f$ and $v$. For that purpose, we adopt a least-square approach. 

If we sample the unit price curves, we obtain a set of $(x,p^{(1)})$ points for each pricing plan. The distance between the actual pricing plan and the two-part model is
\begin{equation}
Q = \sum_{i}\left( \frac{f}{x_{i}}+v-p_{i}^{(1)}\right)^{2}.
\end{equation}
We obtain the best estimates for the two parameters $f$ and $v$ through minimizing $Q$, i.e., zeroing the two derivatives $\partial Q/\partial f$ and $\partial Q/\partial v$. This is tantamount to solving the system of two linear equations
\begin{equation}
\begin{split}
\sum_{i} \frac{1}{x_{i}}\left( \frac{f}{x_{i}}+v-p_{i}^{(1)}\right)&=0,\\
\sum_{i} \left( \frac{f}{x_{i}}+v-p_{i}^{(1)}\right)&=0. 
\end{split}
\end{equation}
For a set of $n$ points the resulting estimates are
\begin{equation}
\begin{split}
\hat{f} &= \frac{n\sum_{i}p_{i}^{(1)}/x_{i}-\sum_{i}p_{i}^{(1)}\sum_{i}1/x_{i}}{n\sum_{i}1/x_{i}^{2}-\left( \sum_{i}1/x_{i}\right)^{2}},\\
\hat{v} &= \frac{\sum_{i}p_{i}^{(1)}-\hat{f}\sum_{i}1/x_{i}}{n}. 
\end{split}
\end{equation}

The resulting values for the parameters are reported in Table \ref{table:parhyper}.

\begin{table}
\begin{center}
\begin{tabular}{lcc}
\toprule
Pricing plan & $\hat{f}$ & $\hat{v}$\\
\midrule
\multicolumn{3}{c}{Consumers}\\
Google Drive & 1.298 & 0.0532\\
IDrive  & 3.42 & 0.0217\\
Dropbox & 7.412 & 0.0667\\
SugarSync & 3.231 & 0.089\\
Symform & 8.58 & 0.106\\
Mozy & 4.91 & 0.033\\
\midrule
\multicolumn{3}{c}{Business}\\
Dropbox & 47.449 & 0.0193\\
IDrive & 6.3 & 0.077\\
Crashplan & 9.674 & 0.267\\
Carbonite & 15.23 & 0.038\\
Mozy & 6.60 & 0.401\\
\bottomrule
\end{tabular}
\caption{Parameters of two-part approximation model}
\label{table:parhyper}
\end{center}
\end{table}

\begin{figure}[htbp]
 \begin{minipage}[b]{5.5cm}
   \centering
   \includegraphics[width=5.5cm]{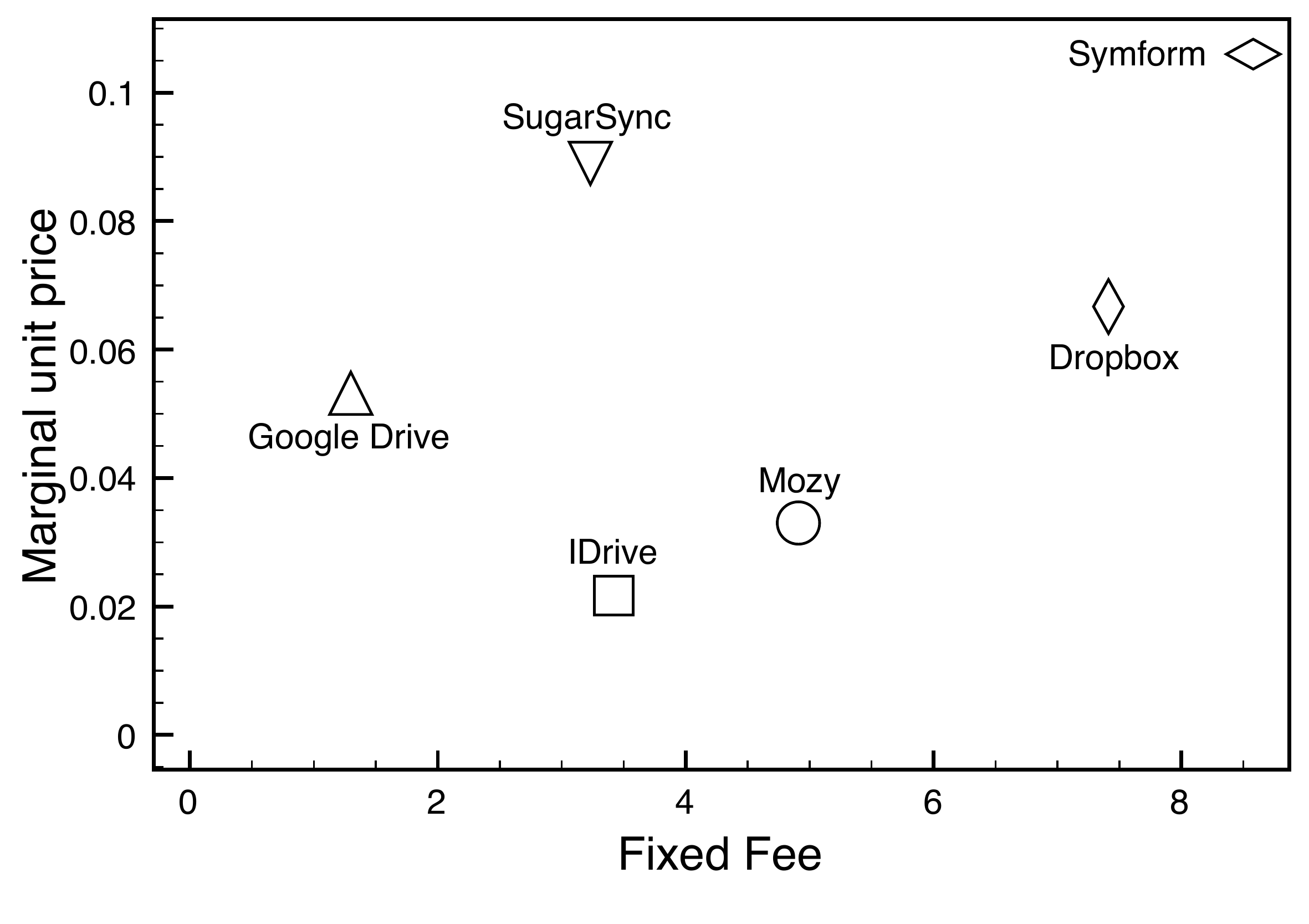}
\caption{Two-part tariff parameters for consumer pricing plans}
\label{fig:cs1}
 \end{minipage}
 \ \hspace{2mm} \hspace{3mm} \
 \begin{minipage}[b]{5.5cm}
  \centering
   \includegraphics[width=5.5cm]{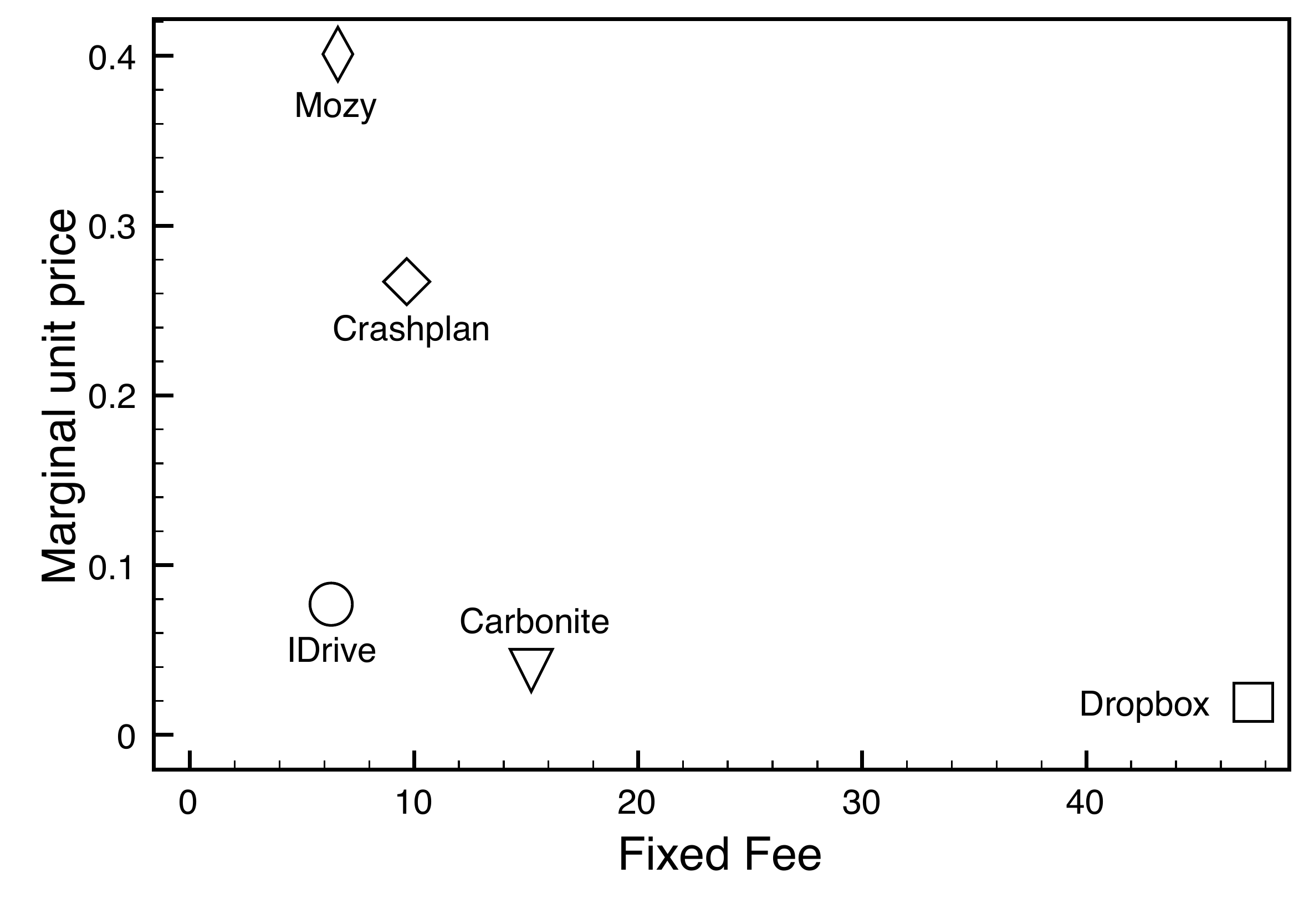}
\caption{Two-part tariff parameters for business pricing plans}
\label{fig:cs2}
 \end{minipage}
\end{figure}

We can now exploit the parameters of the two-part tariff approximations to compare the structure of the pricing plans. We first plot the couple of parameter values obtained for each cloud provider; we report the resulting scatterplot in \figurename~\ref{fig:cs1} and \figurename~\ref{fig:cs2} for consumers and business customers respectively. A price structure is to be considered attractive if both the fixed fee and the variable price per unit are low. On the scatterplots just introduced, the best pricing plans are those represented by points in the bottom left corner (low $\textrm{FF}$ and low $\textrm{VP}$; the worst ones are instead those located in the top right corner. Large differences appear between the pricing plans. In the case of consumers, Symform exhibits the largest values both for the fixed fee and the marginal price; Google Drive provides the lowest fee, but it is Idrive that asks for the lowest marginal price. In the case of business customers, Dropbox provides at the same time the lowest marginal price but the largest fixed fee. 

We can perform an overall comparison among the several pricing plans by using the concept of \textit{Pareto dominance}. The concept of Pareto dominance is of extreme importance in multi-objective optimization. Given a set of objectives, a solution is said to Pareto dominate another if the first is not inferior to the second in all objectives, and, additionally, there is at least one objective where it is better (see chapter 6 of \cite{Ehrgott02}). Pareto dominance has been applied extensively in the context of service tariffing (see, e.g., the works \cite{Miravete96} and \cite{Hoernig07}). Here we take the customer's viewpoint and consider as its objective the minimization of the price whatever the quantity of storage capacity that is leased. In the comparison between two two-part tariff schemes, this objective is reached iff both parameters in Equation (\ref{totcost}) are lower in one of the two scheme instances. We therefore say that a pricing plan dominates another if both its $f$ and its $v$ are lower. In order to identify the most attractive pricing plan, we can therefore eliminate the dominated ones. 

If we look at the consumer pricing plans shown in \figurename~\ref{fig:cs1}, we see that there is not a single dominant plan, but Symform is dominated by every other provider. Both SugarSync and Dropbox are dominated by Google Drive and can therefore be removed from the competition. Mozy (as well as Dropbox) is instead dominated by IDrive. In the end, the best two pricing plans are those offered by Google Drive and IDrive, neither of which dominates the other. In the case of business pricing plans, Crashplan and Mozy are dominated by IDrive, but IDrive does not dominate either Carbonite or Dropbox (though the latter has a very high fixed fee). We end up with three best competitors: IDrive, Carbonite, and Dropbox. 

\section{Conclusions}
We have surveyed the cloud storage packages offered by major providers and put them on a level ground by computing the unit price for each of them over the whole range of capacity values. With the notable exception of Amazon (which follows a block-declining pricing scheme), all the providers adopt a bundling policy. We have compared the pricing plans both on a pointwise basis and on the overall (through a two-part tariff approximation, which uncovers the fixed fee and the marginal price of each pricing plan). Through the first analysis, we determine the cheapest pricing plan for each capacity range. Through the two-part tariff approximation, we apply a Pareto dominance analysis to identify dominated pricing plans, which can be removed from the shortlist of providers from which to choose. In both analyses, a limited number of providers emerge to be considered as prospective providers on the basis of price only.

\end{document}